# Elevating Software Quality in Agile Environments: The Role of Testing Professionals in Unit Testing

Lucas Neves
Cox Automotive Inc.
Recife, PE, Brazil
lucas.macena@coxautoinc.com

Oscar Campos
Cox Automotive Inc.
Recife, PE, Brazil
oscar.campos@coxautoinc.com

Robson Santos
UNINASSAU
Triunfo, PE, Brazil
robsonrtss@gmail.com

Cleyton Magalhaes
Universidade Federal Rural de Pernambuco
Recife, PE, Brazil
cleyton.vanut@ufrpe.br

Italo Santos
Northern Arizona University
Flagstaff, AZ, US
ids37@nau.edu

Ronnie de Souza Santos
University of Calgary
Calgary, AB, Canada
ronnie.desouzasantos@ucalgary.ca

*Abstract*—Testing is an essential quality activity in the software development process. Usually, a software system is tested on several levels, starting with unit testing that checks the smallest parts of the code until acceptance testing, which is focused on the validations with the end-user. Historically, unit testing has been the domain of developers, who are responsible for ensuring the accuracy of their code. However, in agile environments, testing professionals play an integral role in various quality improvement initiatives throughout each development cycle. This paper explores the participation of test engineers in unit testing within an industrial context, employing a survey-based research methodology. Our findings demonstrate that testing professionals have the potential to strengthen unit testing by collaborating with developers to craft thorough test cases and fostering a culture of mutual learning and cooperation, ultimately contributing to increasing the overall quality of software projects.

## I. INTRODUCTION

Software testing is crucial in ensuring software meets user expectations by validating planned features, behaviors, and quality standards. In this context, unit testing focuses on checking small software units—functions or methods—to verify their isolated accuracy and functionality. This approach proactively mitigates the risk of undetected faults progressing into later development phases [1]–[5].

Historically, unit testing has been the responsibility of software developers—programmers or software engineers—to ensure their code integrity before integration [6]–[10]. These tests aim for simplicity, isolation, automation, repeatability, and comprehensive coverage to assure system quality [11], [12]. However, in contemporary agile environments, testing professionals are increasingly involved throughout development to reinforce quality principles and support automation processes [13]. This shift acknowledges their integral role in fostering quality across the project lifecycle.

In this sense, considering the work of testing professionals (e.g., test engineers, test analysts, and quality assurance engineers) in agile environments and the automated-driven nature of unit testing, this study aims to understand how testing professionals contribute to unit testing within the industrial context, guided by the following research question:

*RQ: How do testing professionals contribute to the effectiveness and efficiency of unit testing within software development processes?*

Our study is organized as follows. In Section II, we describe our method. In Section III, we present our results, which are discussed in Section IV. Finally, Section V summarizes the contributions of this study.

## II. METHOD

Most research on unit testing and test automation has traditionally taken a highly technical perspective, primarily concentrating on methods and tools. In our study, we opted to adopt a perspective that centers on the human aspects of the testing activity, and instead of solely emphasizing technical aspects, our focus was on understanding how different professionals engage in and contribute to the unit testing process, particularly within the agile framework.

We conducted a cross-sectional industrial survey [14] following well-established guidelines in software engineering [15] to explore the experience of testing professionals with unit testing. In this process, we developed a questionnaire targeting test engineers, test analysts, and quality assurance engineers to explore their involvement in unit testing within software development. This survey covered aspects like their familiarity with unit testing, practical engagement, benefits of collaboration between developers and testers, and reasons behind the limited historical participation of test engineers in unit testing.

The questionnaire comprised a mix of multiple-choice and open-ended questions, aiming to gather diverse insights. Additionally, the questionnaire incorporated inquiries about participants' backgrounds to enhance the depth of analysis. After formulating the questions, a pilot questionnaire was validated by two test engineers who were not part of the sample and suggested the wording and the sequence of how the questions should appear to participants. The final questionnaire comprised 14 questions, presented in Table I.

TABLE I
SURVEY QUESTIONNAIRE

| |
|---|
| 1. This research aims to explore the understanding of unit testing (concepts and practice) among software testing professionals, along with their experience working with this type of test. This survey is COMPLETELY ANONYMOUS. Answering the questionnaire will take up to 5 minutes. Do you agree to participate?<br>( ) Yes |
| 2. Which gender do you identify with?<br>3. In what country is located the company you are working for?<br>4. What is your highest educational level?<br>5. How long have you been working with software testing?<br>6. Do you have any testing certification?<br>7. What is your job level? |
| 8. What is your theoretical understanding of unit testing (e.g., based on what you studied)?<br>( ) I know a lot about it<br>( ) I have a fair amount of understanding about it<br>( ) I have some understanding about it<br>( ) I have little understanding about it<br>( ) I know nothing about it |
| 9. What is your practical understanding of unit testing (e.g., based on your work with it)?<br>( ) I know a lot about it<br>( ) I have a fair amount of understanding about it<br>( ) I have some understanding about it<br>( ) I have little understanding about it<br>( ) I know nothing about it |
| 10. [OPTIONAL] Please, describe what you know about unit testing.<br>11. [OPTIONAL] If you ever work with unit testing, please describe your experience. |
| 12. Should testing professionals participate in unit testing activities?<br>( ) Yes. It is essential<br>( ) Yes. But it is not essential<br>( ) Only in some particular situations<br>( ) No. This is part of the developer's work |
| 13. [OPTIONAL] Comment on your answer above.<br>14. How software projects can benefit from having testing professionals working alongside developers in unit testing?<br>15. What are the roadblocks that make it difficult for testing professionals to work alongside developers in unit testing? |

We then applied two strategies to collect data from testing professionals: convenience sampling and snowballing [16]. We used convenience sampling to share the questionnaire using our extensive network of software testing professionals who could participate in the study. Further, we used snowballing by asking participants from the convenience sample to forward the questionnaire to colleagues and co-workers who could collaborate with the study.

Concerning post-data collection, we used descriptive statistics [17] to systematically summarize participants' answers, uncovering distribution patterns and frequencies. This method provided an insightful overview of our quantitative data, highlighting key trends within the sample. Furthermore, we employed thematic analysis [18] for open-ended responses, which allowed us to extract recurring themes and intricate details from qualitative data. This approach enriched our understanding of participants' experiences, offering nuanced insights and enhancing the qualitative aspects of the study.

## III. FINDINGS

Our study collected 48 responses from a diverse group of testing professionals, including 17 women and 31 men. Among the participants, 19 individuals had specialized training in software testing, holding advanced degrees in the field (e.g., post-baccalaureate and Master's degrees), thereby enhancing the depth of expertise among the participants. Table II provides a comprehensive overview of their backgrounds. Most participants (79%) had three to five years of industry experience, with 3% having over five years. Additionally, 42% held Senior or Principal positions as test engineers, highlighting the considerable expertise within the sample. The participants come from various professional backgrounds and represent companies from ten different countries (including Argentina, Brazil, Canada, France, Germany, Ireland, Mexico, Spain, Uruguay, and the US), often engaged in software projects with international clients, indicating familiarity with globally recognized software development practices.

TABLE II
DEMOGRAPHICS

| Participants Profile | | |
|---|---|---|
| Gender | Male | 31 individuals |
| | Female | 17 individuals |
| Educational Level | High-School | 4 individuals |
| | Bachelor's degree | 25 individuals |
| | Post-baccalaureate | 13 individuals |
| | Master's degree | 6 individuals |
| Job Level | Trainee | 2 individuals |
| | Junior | 11 individuals |
| | Mid-level | 15 individuals |
| | Senior | 13 individuals |
| | Principal | 7 individuals |
| Testing Experience | 0-1 Years | 5 individuals |
| | 2-4 Years | 5 individuals |
| | 3-5 Years | 21 individuals |
| | 5+ Years | 17 individuals |
| Testing Certification | No | 40 individuals |
| | Yes | 8 individuals |

### A. Knowledge on Unit Testing

Our sample demonstrates a noteworthy distinction between theoretical comprehension and practical experience in unit testing. A substantial 83% of participants possess theoretical knowledge, typically gained through academic coursework or educational sources. However, when it comes to real-world applications, only 1% reported practical experience with unit testing. This disparity underscores a gap between knowledge and hands-on proficiency concerning unit testing, indicating a potential area for further skill development and practical training in agile environments to bridge the divide between theory and practice. Table III summarizes these results and presents the percentage of answers in our sample.

When considering the experience of our sample participants in developing unit tests, the professionals in our study shared their practical insights and experiences employing the following strategies: structured-based testing, path-based testing, and requirement-based testing. Structured-based testing involves creating unit tests for each code statement and meticulously examining conditions and loops to ensure comprehensive coverage of the codebase. Path-based testing requires crafting tests for all potential code execution paths, thoroughly inspecting each conditional structure to cover all possible routes, significantly enhancing test coverage and code resilience. Requirement-based testing aligns unit testing with software specifications, determining test numbers based on typical scenarios and ensuring comprehensive coverage in line with defined requirements.

TABLE III
KNOWLEDGE ON UNIT TESTING

| Level of Knowledge | Theoretical | Practical |
| --- | --- | --- |
| I know a lot about it | 6% | 2% |
| I have a fair amount of understanding about it | 48% | 8% |
| I have some understanding about it | 35% | 2% |
| I have little understanding about it | 11% | 88% |
| I know nothing about it | 0% | 0% |

Finally, regarding their understanding of unit testing, the professionals in our sample provided insights about their experience with unit testing in various contexts. These experiences spanned from its execution within general agile methodologies like Scrum, its integration into processes employing test-driven development (TDD), and its utilization in the context of mutation testing. Their insights shed light on the versatility and adaptability of testing professionals in supporting unit testing practices across different software development approaches.

### B. Involving Testing Professionals in Unit Testing

A substantial portion of our participants recognize the significant role that testing professionals can play in collaborating with developers during the implementation of unit tests. Notably, 54% of respondents consider this collaboration as a fundamental aspect, emphasizing the pivotal contribution of testing professionals in ensuring the effectiveness of unit testing. Additionally, 35% acknowledge the importance, albeit not indispensability, of such collaboration. This perspective highlights the acknowledged synergy between testing professionals and developers in elevating the quality of unit testing practices.

In contrast, a smaller faction, constituting 8% of the sample, believes that test engineers should only be involved in scenarios where their presence is critically necessary. Meanwhile, a mere 2% assert that unit testing should exclusively remain the responsibility of developers. These differing opinions within the sample demonstrate the diversity of perspectives on the role of testing professionals in unit testing practices. Nonetheless, there remains a prevailing sentiment that testing professionals possess valuable expertise to significantly enhance the quality of unit testing, emphasizing their indispensable contribution to improving unit testing practices.

### C. Pros and Cons of Involving Testing Professionals in Unit Testing

In our exploration of involving testing professionals in unit testing, participants highlighted key advantages and challenges. They discussed that test engineers play a crucial role in enhancing test coverage, thereby strengthening the overall robustness of the development process. Their expertise in testing techniques not commonly familiar to developers fosters knowledge exchange, elevating project quality and enabling early detection of failures in initial development stages.

However, our participants also emphasized significant challenges hindering the effective mobilization of test engineers for unit testing in software projects. These challenges include a lack of practical knowledge regarding unit testing approaches, managerial decisions that discourage test engineer involvement due to insufficient planning for quality activities, time constraints as professionals are engaged in multiple testing responsibilities, and a notable lack of enthusiasm among developers for unit testing. Despite acknowledging the benefits, these obstacles hinder the seamless collaboration between testing professionals and developers in unit testing efforts.

## IV. DISCUSSION

Our research findings challenge the conventional academic viewpoint that unit testing should be the sole responsibility of developers [11], [12]. In practice, unit testing demands a specific skill set, commonly found in testing professionals but not always expected from developers, such as knowledge of testing techniques and test coverage [19]–[21]. Below, we present some implications of our study.

### A. Implications

The collaboration between developers and testers has the potential to yield significant outcomes within unit testing. Drawing from our study's insights, it is evident that testing professionals have a substantial theoretical knowledge base in this area (as indicated in Section III-A). Consequently, we posit that testing professionals are poised to:

- Collaborate with developers to plan unit tests, ensuring their alignment with the overall testing strategies and quality objectives of the project, hence guaranteeing that unit tests are not only comprehensive but also effective in achieving the project's goals;
- Assist in designing unit tests by helping developers create comprehensive test cases that encompass a wide array of scenarios, including intricate edge cases, thereby enhancing the efficacy in identifying nuanced issues and providing extensive code coverage;
- Provide valuable input regarding the selection of test data for unit tests to ensure it accurately reflects real-world scenarios, thereby enhancing the authenticity of unit tests in identifying potential issues and validating code behavior within practical contexts;
- Take an active role in refining the unit testing process by integrating best practices from the domain of software

quality, thus fostering a culture of continuous improvement and ensuring its alignment with industry standards and evolving quality benchmarks;
- Share their testing expertise with developers and support a collaborative learning environment, hence facilitating a context where skills, insights, and best practices flow freely between testing professionals and developers in agile environments.

While software testing professionals possess significant theoretical knowledge in unit testing, their practical experience is often limited due to a lack of opportunities. Hence, to address the above-cited recommendations, software project managers must develop some strategies, including: a) allocating time for software testers to participate in unit testing activities; educating developers about the importance of unit testing and the benefits of collaboration with testing professionals; c) motivating the team to actively engage in these endeavors.

*B. Limitations*

At this stage, it is important to acknowledge certain limitations in our study, particularly regarding potential threats to validity. Our study primarily focused on software testing professionals, including software testers, test engineers, test analysts, and quality assurance specialists. However, we recognize that developers play the most significant role in unit testing today. Therefore, exploring their perspectives on the subject can improve the applicability of our findings. Additionally, it is worth reporting that our sample size does not permit broad generalizations. However, we contend that our findings offer valuable insights to practitioners, emphasizing the benefits of collaborations between test engineers and developers in enhancing unit tests and the necessary actions to facilitate such collaborative efforts.

## V. CONCLUSION

In our study, we explored the potential contributions of software testing professionals to unit testing. Leveraging insights drawn from the experiences of industry practitioners, our findings strongly advocate for collaboration between developers and test engineers in the context of unit testing despite the traditional assignment of these responsibilities solely to developers. Testing professionals with a solid theoretical foundation possess the knowledge and expertise necessary to elevate the quality of unit testing. Their familiarity with various quality techniques and approaches positions them as relevant participants in designing comprehensive test cases, automating unit tests, providing insights on test data, and promoting a culture of mutual learning.

In this sense, we can conclude that by bridging the historical gap between certain testing and programming activities, we are supporting more robust and effective quality processes, ultimately enhancing the success of software development projects. To the best of our knowledge, this study stands as the first empirical research to explore the role of test engineers in unit testing within an industrial setting. Our findings pave the way for future research, as numerous aspects of software testing and test automation in agile development could benefit from a similar investigation.